\documentclass[letterpaper]{article}
\usepackage{prepr}

\usepackage[shortcuts]{extdash}
\usepackage{longtable}
\usepackage[framemethod=TikZ]{mdframed}

\newmdenv[%
    backgroundcolor=black!5,
    frametitlebackgroundcolor=black!10,
    roundcorner=2pt,
    linecolor=black!60,
    linewidth=1.1,
    font=\footnotesize,
    skipabove=\topskip-5pt,
    innertopmargin=\topskip,
    splittopskip=\topskip,
    frametitlerule=true,
    nobreak=true,
    ntheorem=false,
    innertopmargin=5pt,
    skipbelow=3pt,
    innerbottommargin=5pt,
]{reqtext}
\newenvironment{reqbox}[1][]
    {
        \begin{reqtext}[  frametitle={\sffamily~#1}]
    }
    {\end{reqtext}}

\title{Requirements Engineering for a Web-based Research, Technology \& Innovation Monitoring Tool}

\author{
\PREPauthor{Alexandra Mazak-Huemer}{Austrian Council for Sciences, Technology, and Innovation (FORWIT), Vienna Austria}{amh@forwit.at}
\PREPauthor{Christian Huemer}{Business Informatics Group, TU Wien, Vienna Austria}{christian.huemer@tuwien.ac.at}
\PREPauthor{Michael Vierhauser}{University of Innsbruck, Department of Computer Science}{Michael.Vierhauser@uibk.ac.at}
\PREPauthor{Jürgen Janger}{Austrian Institute of Economic Research, WIFO}{juergen.janger@wifo.ac.at}
\vspace{0.6cm}
}

\setvenue{\textbf{European Commission: Joint Research Centre}}
\setvenueone{\textbf{European Commission: Joint Research Centre}}
\setyear{2024}
\setdoi{https://publications.jrc.ec.europa.eu/repository/handle/JRC139508}

\begin{document}
\pagestyle{custom}
\maketitle
% \newacronym{ID}{SHORT}{LONG}

% General terms without Glossary entry
\newacronym{cdlsqi}{CDL-SQI}{Christian Doppler Laboratory for Security and Quality Improvement in the Production System Lifecycle}
\newacronym{wp}{WP}{Work Package}

% Terms with missing Glossary entry in Glossary farm
\newacronym{mvmf}{MvMF}{Multi-view Modeling Framework}
\newacronym{tmvm}{TMvM}{Traceable Multi-view Modeling}
\newacronym{mvcm}{MvCM}{Multi-view Change Management}
\newacronym{se}{SE}{Software Engineering}
\newacronym{td}{TD}{Technical Debt}
\newacronym{edalis}{EdaLIS}{Engineering Data Logistics Information System}
\newacronym{cc}{CC}{Common Concept}

% Terms with Glossary entry
\newacronym{cpss}{CPSS}{Cyber-Physical Social System}
\newacronym{cis}{CIS}{Collective Intelligence System}
\newacronym{bft}{BFT}{Byzantine Fault Tolerance}
\newacronym{edkb}{EDKB}{Engineering Data and Knowledge Base}
\newacronym{sut}{SuT}{System under Test}
\newacronym{pse}{PSE}{Production Systems Engineering}
\newacronym{fpd}{FPD}{Formalised Process Description}
\newacronym{ast}{AST}{Abstract Syntax Tree}
\newacronym[longplural={Equivalence Classes}]{ec}{EC}{Equivalence Class}
\newacronym{ecp}{ECP}{Equivalence Class Partitioning}
\newacronym{ppr}{PPR}{Product, Process \& Resource}
\newacronym{dsl}{DSL}{Domain-Specific Language}
\newacronym{pprdsl}{PPR-DSL}{Product, Process \& Resource Domain-specific Language}
\newacronym{ccg}{CCG}{Common Concepts Glossary}
\newacronym{cpps}{CPPS}{Cyber-Physical Production System}
\newacronym{cps}{CPS}{Cyber-Physical System}
\newacronym{bdd}{BDD}{Behavior-Driven Development}
\newacronym{qa}{QA}{Quality Assurance}
\newacronym{pp+r}{PP+R}{Product, Process plus Resource}
\newacronym{cit}{CIT}{Continuous Integration and Testing}
\newacronym{aml}{AML}{AutomationML}
\newacronym{hc}{HC}{Human Computation}
\newacronym{adtree}{ADTree}{Attack--Defense Tree}
\newacronym{dfd}{DFD}{Data Flow Diagram}
\newacronym{iot}{IoT}{Internet of Things}
\newacronym{pprknowledge}{PPRK}{PPR Knowledge}
\newacronym{taf}{TAF}{Test Automation Framework}
\newacronym{ta}{TA}{Test Automation}
\newacronym{hbi}{HBI}{Human-based Inspection}
\newacronym{mde}{MDE}{Model-Driven Engineering}
\newacronym{twin}{DT}{Digital Twin}
\newacronym{dlt}{DLT}{Distributed Ledger Technology}

\newacronym{fmea}{FMEA}{Failure Mode and Effects Analysis}
\newacronym{opcua}{OPC~UA}{OPC Unified Architecture}
\newacronym{f2p}{FMEA+PPR}{FMEA-linked-to-PPR assets}
\newacronym{cm}{CM}{Configuration Management}
\newacronym{pad}{PAD}{Production Asset Directory}
\newacronym{pan}{PAN}{Production Asset Network}
\newacronym{i40}{I4.0}{Industry 4.0}

\newacronym{dspqa}{DS-PQA}{Digital Shadow for Production Quality Analysis}

\newacronym{cen}{CEN}{Cause-and-Effect Network}
\newacronym{qdg}{QDG}{Quality Dependency Graph}

\newacronym{fps}{FPS}{FMEA+PPR+Skills}
\newacronym{f2o}{F2O}{FMEA-to-Operation}

\newacronym{mmr}{MMR}{Multi-Model Reviewing}
\newacronym{fpi}{FPI}{FMEA-linked-to-PPR Asset Issue Analysis}

\newacronym{rabpc}{RABPC}{Risk-Aware Bin Picking Configuration}
\newacronym{ibp}{IBP}{Industrial Bin Picking}
\newacronym{ibpa}{IBPA}{Industrial Bin Picking Application}

\newacronym{pqs}{PQS}{Production Quality+Security}
\newacronym{dt}{DT}{Digital Twin}

\newacronym{i4an}{I4AN}{Industry 4.0 Asset Network}

\newacronym{mdeg}{MDEG}{Multi-Disciplinary Engineering Graph}
\newacronym{sql}{SQL}{Structured Query Language}
\newacronym{mdre}{MDRE}{Model Design and Review Editor}

\newacronym{ptsv}{PTSV}{Production Test Scenario Validation}
\newacronym{pts}{PTS}{Production Test Scenario}
% \newacronym{pts}{PTS}{production test scenario}

\newacronym{pds}{PDS}{Production Data Space}
\newacronym{pprs}{PPRS}{Product, Process, Resource \& Skill}
\newacronym{can}{CAN}{Coordination Asset Network}

\newacronym{aft}{AFT}{Agile Field Test}
\newacronym{is}{IS}{Information System}
\begin{abstract}
  With the increasing significance of Research, Technology, and Innovation (RTI) policies in recent years, the demand for detailed information about the performance of these sectors has surged. Many of the current tools are limited in their application purpose. To address these issues, we introduce a requirements engineering process to identify stakeholders and elicitate requirements to derive a system architecture, for a web-based interactive and open-access RTI system monitoring tool. Based on several core modules, we introduce a multi-tier software architecture of how such a tool is generally implemented from the perspective of software engineers. A cornerstone of this architecture is the user-facing dashboard module. We describe in detail the requirements for this module and additionally illustrate these requirements with the real example of the Austrian RTI Monitor.
\end{abstract}

\glsresetall

\section{Introduction}
\label{sec:intro}
%%%%%%%%%%%%%%%%%%%%%%
% What is an RTI Monitoring System?

With the increasing significance of Research, Technology, and Innovation (RTI) policies in recent years, the demand for detailed information on the performance of these sectors has surged. This heightened interest has led to the consolidation of statistical data, and has further spurred the refinement of monitoring systems across various institutions and programs. Additionally, there is an increasing need for precise, yet comprehensive, information for a variety of different target audiences. Such data then serves as the foundation for evidence-based policy-making, catering to the diverse needs of stakeholders within the RTI community.

Innovation rankings, such as the European Innovation Scoreboard~\cite{EIS} by the European Commission, and the Global Innovation Index~\cite{GII} by the World Intellectual Property Organization use indicators to rank countries by performance. These rankings typically rely on composite indicators that average over individual innovation metrics~\cite{Arnold2024}. Although these tools simplify communication with policymakers~\cite{Grupp2010}, they often obscure the underlying factors that drive performance. As a result, this obscuration frequently leads to only superficial discussions about enhancing countries' rankings~\cite{Arnold2024}. Even more sophisticated, interactive visualizations of these rankings tend to present mere lists of indicators organized by linear production logic, rather than a more immersive way of exploring information. Further drawbacks stem from the fact that underlying factors driving performance often remain overlooked and do not provide a systemic perspective on how various factors influence innovation performance, as presented in~\cite{janger_2020}.

The limitations of current systems do not stem from a lack of information but from the complexity of the subject itself. There is also an overload of data that challenges integration and varies in its utility among different actors, such as ministries, agencies, and research institutes. Moreover, many of these tools are restricted to specific purposes, like annual reports or bespoke analyses, and some do not offer open access~\cite{Arnold2024}.

To address these issues, we create a three-tier software architecture based on an approach from the Software Requirements Engineering field to set up a web-based interactive and open-access RTI system monitoring tool. We base our work on the so-called viewpoint-guided Requirements Engineering process, the VORD model introduced by \cite{kotonya1996requirements}, respectively. In general, the requirements for a software system to be developed are classified into \textit{functional requirements} (i.e., components of a system that software engineers must implement to enable users to accomplish their tasks) and \textit{non-functional requirements} (e.g., set of specifications such as usability, reliability, data integrity of the system to be developed). The software system we focus on aims to cover several core features to enable the monitoring of an RTI system. For instance, it should allow for analysis of strengths and weaknesses of various aspects of an RTI system, e.g., to facilitate the review of impacts of past policies. It should further enhance an understanding of the systemic interdependencies within a system by exploring relationships among different RTI indicators. Additionally, the system should allow for comparisons between a single country's innovation system and those of others, e.g., comparing a selected country with the innovation leaders defined by the European Innovation Scoreboard~\cite{EIS}.

%% updated structure of the paper
The paper is structured as follows. 
In \citesec{RE_Process}, we describe a lightweight, viewpoint-guided Requirements Engineering process to elicitate the core components and functionalities required for a three-tier software architecture. Based on this, in \citesec{Components}, we present this architecture and subsequently describe in depth its general components. In~\citesec{requirements}, we focus on a particular component, the \textit{Dashboard Module}, on which we present specific requirements from the users' perspective when using the RTI monitoring tool. For better comprehensibility, we refer in gray boxes to a real example -- the Austrian RTI Monitor~\cite{monitor}, currently available in its third version, commissioned by the Austrian Council for Sciences, Technology, and Innovation (FORWIT, former RFTE) and implemented by the Austrian Institute of Economic Research (WIFO) (the first prototype was made online in 2022). We close this paper by giving an outlook in Section~\ref{sec:conclusion}.

\section{A Viewpoint-guided Requirements Engineering Process}\label{sec:RE_Process}
%%%%%%%%%%%%%%%%%%%%%%%%%%%%%%%%%%%%%%

Almost 50 years ago, Fred Brooks argued that``\textit{the hardest single part of building a software system is deciding precisely what to build}''~\cite{brooks2021mythical}.
\textit{Requirements Engineering} inherently demands the active involvement and interaction of success-critical stakeholders, each bringing unique perspectives and objectives. The challenge resides in facilitating a process that is both accessible to non-technical participants and capable of producing technical deliverables that meet the stringent needs of Requirements Engineering~\cite{pohl1996requirements}. 

Requirements Engineering activities generally take place in the early phases of a software development lifecycle, starting with the need to understand, collect, and document the desired outcome, as well as the potential impediments hampering the successful system realization. The main objective of the requirements engineering process is to provide a model of what is needed in a clear, consistent, precise, and unambiguous manner~\cite{kotonya1996requirements}. In case of an online RTI monitoring system, with a broad and diverse user base, the key functionalities are to collect, prepare, store, analyze, and visualize a plethora of data gathered from reliable and trustworthy data sources from all over the world. To successfully identify all relevant requirements for the different parts of such a system stakeholders who are to work with this system and stakeholders who are to develop and implement it need to be identified.

%%maybe remove this paragraph.. too much stakeholder stuff..
% It is important to note that stakeholders do not solely comprise the end users who are interested in using the system, but also include any group or individual who can affect or are affected by the system. This includes, for example, developers responsible for design and development; people responsible for its sale or purchase, and personnel responsible for introduction and maintenance. For the majority of cases, neither the requirements engineer is an expert in the application domain, nor all stakeholders possess the necessary technical background to formally specify clearly and consistent requirements based on their limited familiarity with the entire system, technical constraints, or other limitations.

From the \textit{realization perspective}, three core stakeholder groups are involved. First, the ``customers'' interested in establishing such a monitor; second, the ``domain experts'' who provide complex analyses and interpretation of the raw data; and third the ``end users'' looking at the prepared data to support their decision-making. They all have different domain knowledge and roles, and therefore, are interested in using the tool for their specific purposes in their daily business. In a research, technology, and innovation context, the customers are governmental or national entities; the domain experts are often economists, and the end-users are, for instance, administrative officials and politicians.

For this reason, in Requirements Engineering, several strategies, such as workshops, surveys, and introspection techniques have been developed, aiding in the identification and collection of requirements for software systems in a wide range of application fields (like discussed in \cite{2013analysis}). A specific approach that is particularly helpful for incorporating different angles on a system to be implemented are \textit{Requirements Viewpoints} to efficiently categorize them. 
Initially described by~\cite{kotonya1996requirements} as \textit{VORD (Viewpoint-Oriented Requirements Definition)}, viewpoints allow structuring requirements around different types/classes of stakeholders that directly or indirectly interact with the different parts of the system. Viewpoints, can reside on different levels and cover different aspects of system development. For example, a system engineering viewpoint enables to focus on technical aspects of a system such as security, usability, or system performance to be considered in the system architecture.
By concentrating on more stakeholder-oriented perspectives, viewpoints can also highlight specific core aspects of a system.  For example, the end-user viewpoint emphasizes user interfaces, the business logic viewpoint addresses the system's core functionality, and the data-related viewpoint focuses on the management of underlying data functions.
\begin{figure}[t!]
    \centering
    \includegraphics[width=1\linewidth]{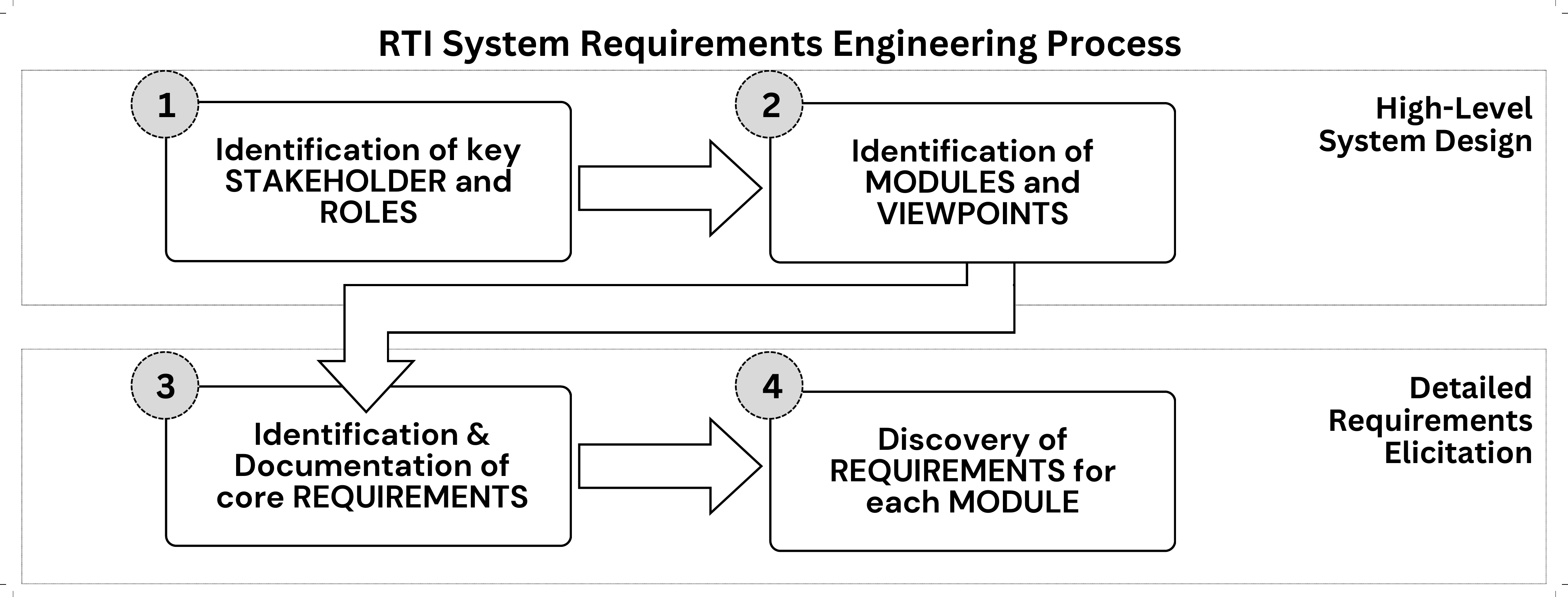}
    \caption{A Workflow-oriented View on the Core Steps of a lightweight Viewpoint-guided Requirements Engineering Process.}
    \label{fig:RE_workflow}
    \vspace{-0.5cm}
\end{figure}

In addition to the specific viewpoints, a viewpoint-guided Requirements Engineering process commonly includes several core activities. First,  {\small\textcircled{\normalsize \texttt{1}}} identifying all involved stakeholders both \textit{direct} -- which ultimately will use and interact with the system -- as well as \textit{indirect} -- who are affected by the system, e.g. via interfaces or modules, but do not directly interact or use the system~\cite{glinz2007stakeholders}.
 This typically involves a broad range of roles, domain experts, software engineers, developers, and business analysts which requires {\small\textcircled{\normalsize \texttt{2}}} different viewpoints to be considered to satisfy all needs and {\small\textcircled{\normalsize \texttt{3}}} identifying core features and the respective requirements, as well as cross-cutting or even conflicting requirements during the process. Ultimately, {\small\textcircled{\normalsize \texttt{4}}} using negotiation techniques, such for example the WinWin methodology introduced by \cite{boehm1998using}, a set of documented, validated, and approved requirements can be established that guide further design and development of the system. This is commonly done in an interactive manner, where multiple cycles of prototyping and refinement lead to an incrementally growing system. 

For this paper, we put a particular emphasis on aspects concerning the establishment of the different viewpoints relevant to specifying the core requirements of a high-level system architecture to realize an RTI monitoring tool. Each selected viewpoint represents its own interests, concerns, or knowledge of a specific group of stakeholders. Similar to requirements, viewpoints can be categorized in different ways. For example, \textit{functional viewpoints} focus on the specific functionalities the system must provide, often aligned with the business goals of the stakeholders. In the case of the RTI monitoring tool, these functional or technical viewpoints guide the development of the high-level architecture (cf. Section~\ref{sec:Components}) and its core components from a software engineer's perspective. \textit{Non-functional viewpoints} on the other hand focus on aspects of system performance or security aspects but also cover end-user-related aspects, for example, the requirement for an easy handling of the tool. This \textit{user viewpoint}, focusing on the dashboard, includes concerns, for example regarding the intuitiveness of the user interface, or learnability (how easy it is for new users to understand different types of visualizations). We specifically leverage these three types of viewpoints, functional, technical, and non-functional for creating and motivating the core architecture of the introduced RTI monitoring system as introduced in the next section.

In the remainder of this article, we demonstrate how such a Requirements Engineering process together with these viewpoints can be utilized first to identify and select key features and modules of an RTI monitoring framework (\citesec{Components}) and how requirements for a specific module (i.e., the Dashboard Module) are collected and documented (\citesec{requirements}).

\section{Components of an RTI Monitoring System}
\label{sec:Components}

% Describe the components of an RTI Monitoring System

To set up an RTI monitoring system, RTI-relevant data must be retrieved from a number of different trustworthy \textit{sources}. This data then needs to be integrated into the RTI monitoring system and stored in its \textit{Data Store}. Subsequently, an \textit{Analytics Engine} can access this data, to carry out the corresponding data processing, aggregation, and analyses. The results must be visualized in the form of user-friendly \textit{Dashboards}. Furthermore, the RTI monitor framework comprises a dedicated Configuration Module, responsible for  (a) data source configuration management and (b) user management. The former determines which data sources are used for the monitoring, and how these data sources are transformed to become accessible to the monitor, whereas the latter is an optional functionality to define who is allowed to perform which analytics, and how the results are visualized. These configurations are specified by an administrator of the monitor via a dedicated user interface. Another optional component is the Notification Module that proactively informs key persons about changes in the RTI system in order to initiate appropriate interventions.

\begin{figure}[t!]
   \centering
   \includegraphics[width=1\linewidth]{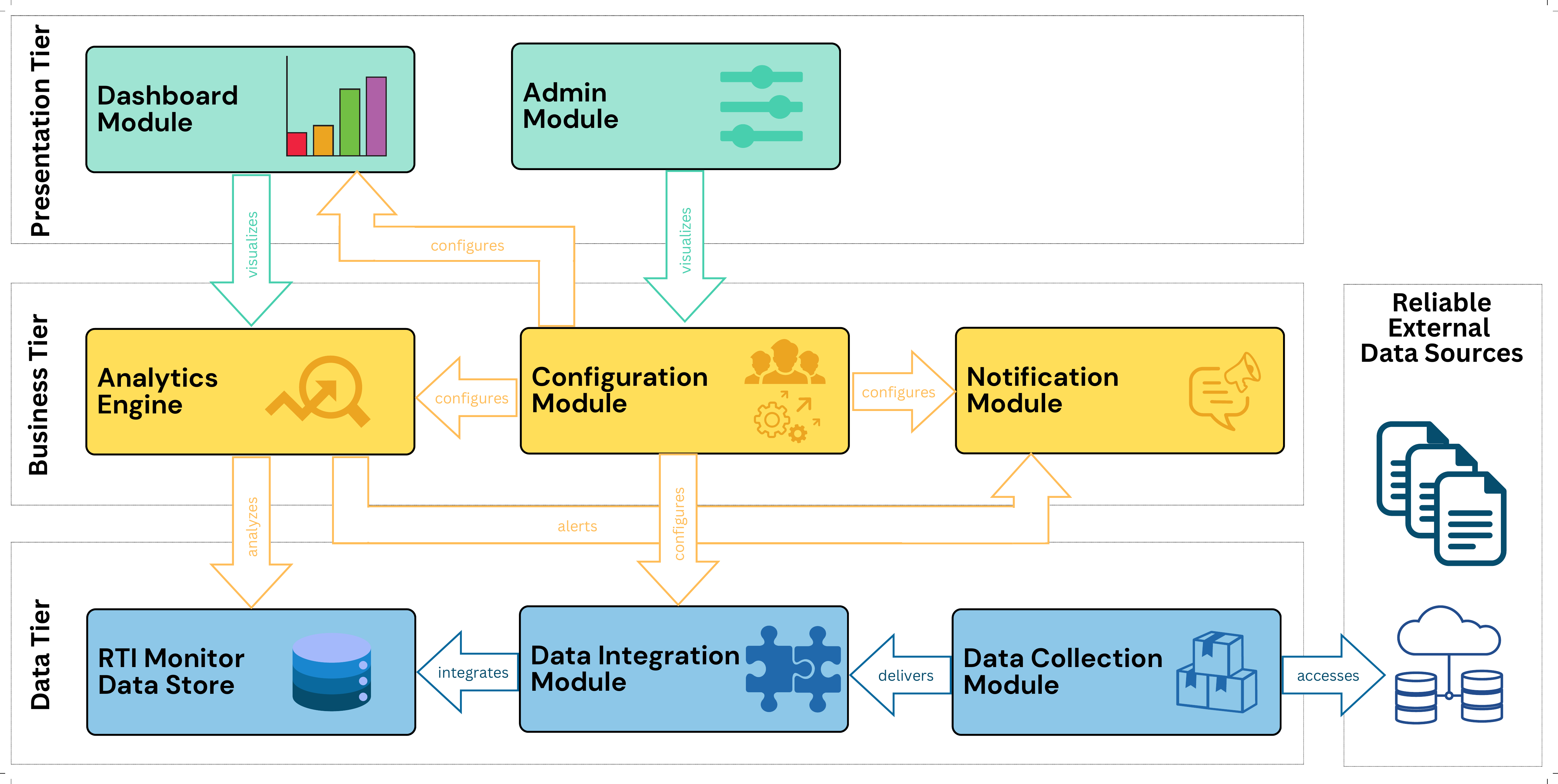}
   \caption{Three-Layer Basic Components Architecture for a Web-based Monitoring Tool.}
%   \caption{RTI Monitoring Software Architecture.}
   \label{fig:architecture}
   \vspace{-0.5cm}
\end{figure}

To provide the core functionality, covering the base requirements of an RTI monitoring system, we derived a high-level, three-tier architecture as depicted in Figure~\ref{fig:architecture}. 
The bottom tier (\textit{Data Tier}) is responsible for collecting information from predefined sources, processing it, and storing it for subsequent analysis.
This data is then leveraged by the \textit{Business Tier}, providing configuration functionality for data and user access, and analytics functionality. 
Finally, the top \textit{Presentation Tier}  provides appropriate visualizations and user-tailored views for end-users and administrators, respectively.
 All the different modules of these three tiers will be discussed in detail in the following subsections.

\subsection{Data Collection Module}

The primary purpose of the \textit{Data Collection Module} is to systematically gather data from various reliable sources, establishing a comprehensive and accurate foundation for the RTI Monitor. It ensures that the RTI Monitor has access to high-quality data, which is essential for making informed decisions. 

Accordingly, the Data Collection Module needs to interface with multiple reliable data sources, that provide online access to data relevant for the calculation of indicators of the RTI system. By establishing these connections, it may either collect data on request by an administrator of the RTI Monitor, or it may continuously acquire relevant data. The latter is crucial for delivering up-to-date insights and allowing stakeholders to respond promptly to emerging issues or opportunities.

A high quality of the collected data is paramount. The module includes mechanisms for data validation, error detection, and correction. This guarantees that only accurate, reliable, and complete data enters the RTI Monitor database, reducing the risk of errors in subsequent analysis and reporting.

\subsection{Data Integration Module}

The \textit{Data Integration Module} is designed to ensure that the diverse data collected from various sources are harmonized, consolidated, and made accessible for analysis and decision-making. This module's primary purpose is to create a cohesive and unified dataset from disparate data inputs. Thereby, it promotes interoperability between the RTI Monitor and external systems. 

In the case of online data access, the Data Collection Module delivers the data from the various sources to the Data Integration Module, which aggregates this data. Aggregating the data means also consolidating the data, so that all relevant information is brought together into a single, coherent dataset.

Since different data sources often exhibit different formats, encodings, and structures, the Data Integration Module transforms this heterogeneous data into a format that is compatible with the RTI Monitor Data Store. This transformation includes data cleansing, normalization, and enrichment for the purpose of consistency and usability. For this purpose, it performs data mappings to align data fields from different sources. This process involves identifying corresponding fields in different datasets and mapping them to a common schema. It identifies and removes duplicate entries to avoid redundancy.

Furthermore, the Data Integration Module implements rigorous data quality checks to verify the accuracy, completeness, and reliability of the integrated data. This includes validation rules, error detection mechanisms, and processes for correcting identified issues.

\subsection{RTI Monitor Data Store}

The \textit{RTI Monitor Data Store} provides all the functionality to store, manage, and facilitate access to the integrated data collected from various sources. Its primary purpose is secure, organized, and efficient management of all data.

\begin{figure}[t!]
    \centering
    \includegraphics[width=1\linewidth]{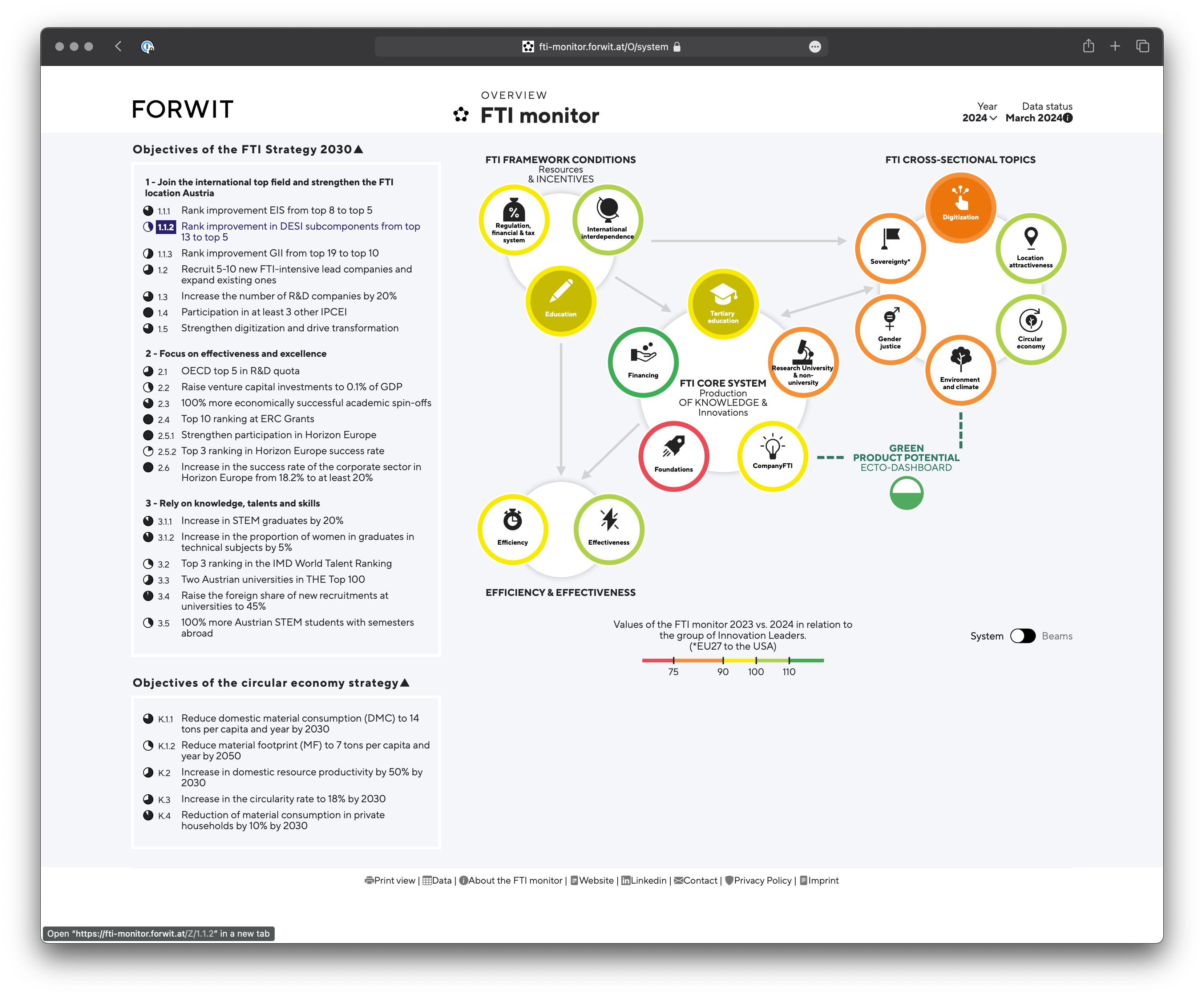}
    \caption{Level 1 of the Austrian RTI Monitor.}
    \label{fig:RTI-Level-1}
    \vspace{-0.99cm}
\end{figure}

Consequently, the Data Store represents a unified storage for all relevant monitoring data. Thereby, all relevant information is stored in a single, easily accessible location, which may be either on-premise or in the cloud. It facilitates efficient data retrieval for users and other system components, enabling quick access to the information needed by the analytics engine of the RTI Monitor.  

The Data Store provides the integrity and security of stored data through access controls and comes with regular backups. It optimizes data performance by using efficient storage and retrieval techniques so that the system can handle complex queries and large datasets without compromising speed or accuracy.

\subsection{Analytics Engine}

The \textit{Analytics Engine} is the core component of the RTI Monitor, designed to transform input data into meaningful insights that drive an informed decision-making process. Its primary purpose is to analyze the collected data stored in the RTI Monitor Data Store and integrate them within the RTI Monitor system.
It employs sophisticated statistical, and complex computational techniques to extract actionable insights, identify trends, and uncover patterns within the data. However, it should be noted that there is a difference between descriptive inference (like we will discuss for the RTI Monitor) and causal inference.

It calculates indicators related to RTI for the purpose of evaluating the performance and effectiveness of, for example, research policies and programs. By analyzing historical and current data, the Analytics Engine identifies trends and supports uncovering patterns in research outputs, funding utilization, and compliance behavior. This trend analysis helps predict future developments and informs strategic planning.

The module manages complex relationships between different data entities. For example, linking research projects to funding details, compliance records, and publication outputs. This relational handling is vital for comprehensive analysis and reporting.

\subsection{Dashboard Module}

The \textit{Dashboard Module} is the user interface of the RTI Monitor, designed to present complex data insights in a user-friendly and accessible manner. Its primary purpose is to visualize data, generate reports, and supply stakeholders with actionable information to monitor research activities, assess policy compliance, and make informed decisions. The module converts the results of the Analysis Engine into intuitive visual formats such as charts and graphs. These visualizations make it easier for users to understand trends, patterns, and key metrics at a glance.

The module receives the results of the Analytics Engine to visualize the indicators of the RTI system. Performance tracking helps stakeholders in assessing progress toward strategic goals and identifying areas for improvement. It allows a comparison of the RTI system of the target region with other relevant regions. Furthermore, it provides access to and analyses of historical data, showcasing a comprehensive view of trends over time. The dashboard includes interactive features such as drill-downs and filters, allowing users to explore data in greater detail and from different perspectives. These features enhance the usability and flexibility of the tool.

In addition, the module also delivers detailed reports that summarize findings, trends, and performance metrics. These reports can be tailored to meet the needs of different audiences, including policymakers, researchers, and administrators. Users can export data and reports in various formats (e.g., PDF, Excel) for further analysis or sharing with other stakeholders. This capability supports collaboration and transparency.

The Dashboard Module significantly enhances the functionality and effectiveness of the RTI Monitor by unveiling valuable insights to stakeholders. By presenting data in an easy-to-understand format, the tool supports data-driven and evidence-based decision-making and strategic planning, leading to better-informed decisions. Visualizing comparisons with other regions and trends helps identify areas of non-compliance and facilitates corrective actions. Monitoring key metrics and performance indicators enables continuous improvement of research policies and outcomes. Clear and accessible information fosters transparency and accountability in research activities and funding.

Overall, the Dashboard Module is essential for transforming raw data into actionable insights guaranteeing effective monitoring, evaluation, and management of research policies and activities. It ensures that stakeholders at all levels have the information they need to drive excellence in research and innovation. We have chosen the Dashboard Module (therefore this area is also colored grey in Figure~\ref{fig:architecture}) to describe the  \textit{Requirements Engineering} process necessary for implementing the user interface for an RTI Monitor, which we describe in-depth on the example of the Austrian RTI Monitor in section \ref{sec:requirements}.

\subsection{Notification System}

The \textit{Notification System} is an optional module of an RTI Monitor. It closes the gap between data analysis and user action by communicating critical information to stakeholders. Its primary purpose is to issue alerts and notifications based on the results of the Analytics Engine, enabling users to take prompt and appropriate actions. 

Thus, the Notification System is implemented to proactively inform users about potential issues or opportunities. The notifications can be delivered through different channels. By means of the Notification System, stakeholders are constantly informed about critical developments, enhancing the overall effectiveness and responsiveness of the Monitor.

\subsection{Configuration Module}

All the modules described so far may be hard-coded which results in a rigid monitoring system. This would imply that adding a new data source for new indicators, updating the format of a data source by an external provider, changing the calculation of certain indicators, or switching to a new visualization of certain indicators all require changes to the code base, and there may be a lot of other example cases. Evidently, hard-coded systems come with high maintenance costs. Accordingly, it is desirable to introduce greater flexibility, enabling modifications and customizations through a configuration module. This module allows for flexible, customizable management of the RTI Monitor without requiring hard-coded changes.

The \textit{Configuration Module} configures the Data Integration Module by specifying which input data, at which location, and in which format, should be included in the RTI Monitoring system. This means the Configuration Module specifies where to find certain input data, and how to access it (note this information must also be forwarded from the Data Integration Module to the Data Collection Module). We already know that the Data Integration Module transforms the input data to a format suitable for import into the RTI Monitor Data Store. Accordingly, the Configuration Module, on the one hand, informs the Data Integration Module about the format of the input data (as defined by the external source) and, on the other hand, specifies the transformation rules to convert the input data to the import format.

Thereby, the monitor becomes scalable and adaptable to handle a growing volume and variety of data. It can accommodate new data sources and adapt to evolving research environments and technological advancements, enabling its long-term relevance and utility.

While much of the data collection is automated today, not all sources are directly available and accessible online. Thus, additional support for manual data entry, for cases where automated integration is not possible needs to be provided. Accordingly, the Configuration Module may further be used to manually enter off-line data sources to be imported into the RTI Monitor Data Store. Thereby, all necessary data, even from less technologically advanced sources, can be included in the monitor.

The configuration of the Analytics Engine is at the heart of the Configuration Module. The Configuration module specifies which data of the Data Store is used to calculate in which way which indicator. In other words, it is used to declaratively specify the functionality of the Analytics Engine. It defines which indicators are calculated by the Analytics Engine and how each of them is calculated. Evidently, the calculation of each indicator must be based on specific data from the Data Store.

The Configuration Module is also used to configure the dashboard and customize the visualizations of the different indicators. It defines on which page of the dashboard which information and which indicators are presented by which visualization types (charts, graphs, lists, etc.). Usually, the Configuration Module offers different kinds of pre-defined template pages. For each page of the dashboard, a template page is selected, and for each area in the template, the information/indicator to be displayed is specified. For each indicator, the visualization type is selected from several predefined types. Furthermore, links between the dashboard pages are to be configured.

\begin{figure}[t!]
    \centering
    \includegraphics[width=1\linewidth]{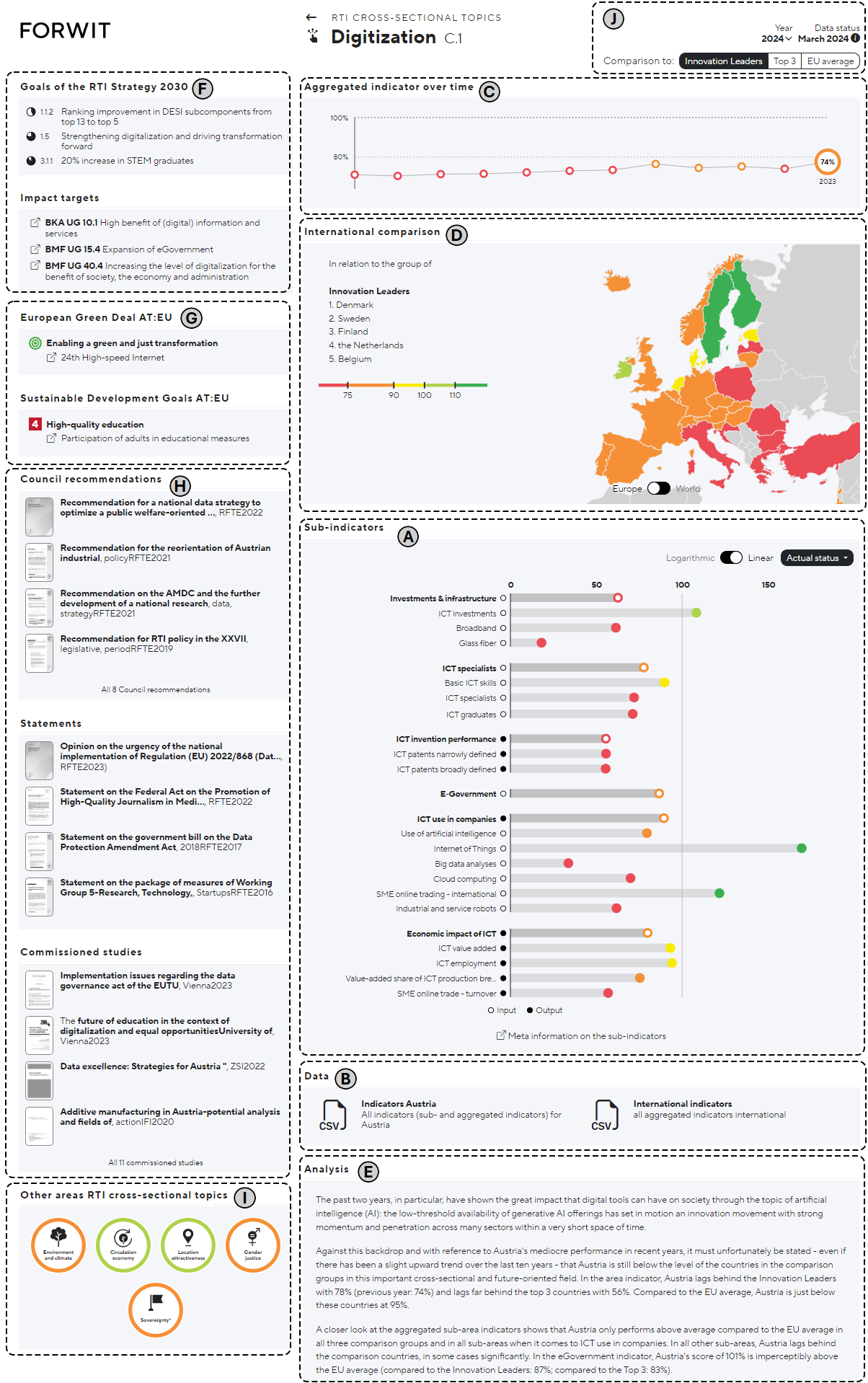}
    \caption{Level 2 of the Austrian RTI Monitor: Analysis of a sub-area}
    \label{fig:RTI-Level-2}
    \vspace{-5mm}
\end{figure}

An optional functionality of the Configuration Module is user management. It is used to define who has access to which dashboard pages and, thus, to which kind of analytics of which indicators. Also, it is used to specify who will receive which kind of notifications by which communication channel. Thereby, users receive only relevant information without being overwhelmed by unnecessary alerts. 

Accordingly, the user management takes care of user access, roles, and permissions within the system. Its primary purpose is to ensure that only authorized personnel can access specific data and functionalities, enhancing security and compliance. It facilitates user administration by adding, modifying, and removing user accounts. It assigns user roles and permissions based on their responsibilities. An activity logging feature may track user activities for accountability and auditing purposes. 

However, it should be noted that a user management function is not always necessary, e.g., in the case of the Austrian RTI Monitor, which is used as an example in section \ref{sec:requirements}, it was decided that all information should be publicly visible to everyone to foster transparency.

Advanced features of a Configuration Module may also be realized at the intersection of user management and data configuration. For example, an advanced feature may allow users to define specific queries and criteria for tailored insights. This flexibility enables stakeholders to obtain the precise information they need for their unique requirements. Another example of an advanced feature is the customization of user-specific dashboards to display the specific data and metrics that are most relevant to their roles and responsibilities. With this customization feature, each stakeholder can focus on the information that matters most to them. The tool can automate the generation and distribution of reports, and thus, stakeholders may receive regular updates without manual intervention. Automated reporting saves time and reduces the risk of errors.

\subsection{Admin Module}

The Admin Module is the user interface for administrators of the RTI Monitor. The administrator uses this interface for the configuration of the RTI Monitor. Evidently, it must provide suitable visualizations and interaction capabilities to specify all configuration features of the Configuration Module as described above.
\section{Requirements for Implementing the Dashboard Module}
\label{sec:requirements}

In the context of governmental or supranational environments, the steering function is of the highest priority. In other words, the monitor must be an easy-to-use, and easily comprehensible tool, showing at a glance the effectiveness of RTI policies in meeting the strategic goals of the government or supranational body.

This, in turn, means that providing an ``easy-to-use'' tool to stakeholders requires considering their needs in steering the RTI system right from the beginning. It is essential to capture and analyze their expectations of the system, as well as their domain knowledge, and transform these insights into an appropriate interaction interface for monitoring the RTI system. The main interaction point in the case of the RTI monitor is the dashboard module represented in the high-level architecture as depicted in Figure~\ref{fig:architecture}.
As part of the aforementioned requirements engineering process, in this section, we discuss (4), the discovery and documentation of the requirements for the dashboard module used by stakeholders to interact with the monitoring tool. 
As a running example,  illustrating requirements of a concrete monitoring system,  we refer to the recently established Austrian RTI-Monitor\footnote{\url{https://fti-monitor.forwit.at}}which was developed and implemented by a team of the Austrian Institute of Economic Research (WIFO) (Arnold et al.~\cite{Arnold2024}) and the Austrian Council for Sciences, Technology, and Innovation FORWIT (former RFTE).

In the following, we, therefore, present examples of how the general requirements(cf. Fig.~\ref{fig:architecture}, Presentation Tier, and Business Tier) were realized in the  Austrian RTI-Monitor~\cite{monitor} (cf. requirements highlighted in the grey boxes).

The requirements elicitation process, including the various diverse stakeholders having different roles, domain knowledge, perspectives, and needs on a monitoring system has revealed to us, that the RTI monitor needs to cover both the \textbf{strategic goals} set by the government or supranational body and the \textbf{RTI policies} intended to meet these goals. Since this typically involves a lot of different RTI policies in place, additional requirements are concerned with the appropriate grouping into different RTI areas, as well as the visualization of the inter-dependencies of these areas. 

In the case of the Austrian RTI monitor, the starting point for the inception of the RTI monitoring system was a hard-copy, annually printed version of an extensive report measuring the performance of research, innovation, and technology development in Austria published by the Austrian Council for Sciences, Technology, and Innovation, based on the work by Janger and Strauss~\cite{janger_2020}, built on the large literature on innovation systems.

On a high level, the main entry/landing page of the dashboard should visualize the various strategy goals and top-level RTI areas in an interrelated manner. 
Interrelated, in this context means, that it should not be simply a listing of goals and areas, but rather intuitively visualize their systemic interactions to underpin which area has which effects on which strategic goals. 
Thereby, systemic correlations become visible with their direct
\begin{reqbox}[Visualizing strategic goals and RTI areas and sub-areas] 

The top RTI areas were identified according to Janger and Strauss~\cite{janger_2020}.
The starting page of the Austrian RTI Monitor is depicted in Figure~\ref{fig:RTI-Level-1}. It shows the identified areas on the right. To visualize their inter-dependencies, these areas are organized as a semantic graph with nodes and edges. The nodes symbolize the RTI areas and sub-areas that describe the RTI system~\cite{Arnold2024}. The edges show the connections among these areas. To measure the innovation performance of Austria's RTI system there are currently 16 sub-areas clustered into 4 main areas: i) Framework Conditions for RTI with its sub-categories: Regulation, Financing and Taxes, Education, International Interdependence; ii) Core RTI System with its sub-areas: Tertiary Education, Academic Research, Corporate RTI, Start-ups, and Financing; iii) RTI Cross-cutting Issues with the sub-areas: Digitization, Environment and Climate, Circular Economy, Location Attractiveness, Gender Equality, and Technological Sovereignty; and iv) the area Impact with its categories: Effectiveness and Efficiency.

In addition, there are the Austrian Federal Government's strategy goals such as the RTI-strategy 2030~\cite{strat2030} and the circular economy strategy~\cite{circ} which are of interest to monitor for different stakeholders from the RTI-community. On the left of the RTI-Monitor, Austria's RTI strategy goals 2030 and the strategic goals for circular economy are displayed. 
\end{reqbox}

%\vspace*{5mm}
 and indirect leverage effects. 
For this purpose, the monitoring tool should offer a start page with the possibility of switching between different perspectives, as well as a mapping of the RTI strategic goals with the RTI system.

%\vspace*{5mm}
\begin{reqbox}
 [Visualizing interrelation between the RTI system and goals]
   In the RTI-Monitor, there is a mapping algorithm developed to interrelate the goals to the system. For this purpose, a functionality was implemented to highlight all of the sub-areas of the system and the specific goals related to each other by using the cursor when moving the mouse over (whether one of the goals or a sub-area was selected). This allocation is based on the performance pinpointing with those indicators that measure the performance of the respective area. The allocation is bidirectional -- beginning with the strategic RTI goals, reference is made to the respective RTI subsystems and their areas and vice versa depending on the users' point of view.
   
   For example, when selecting the RTI strategic goal of "Ranking improvement in DESI sub-components from top 13 to top 5" (see Figure~\ref{fig:RTI-Level-1} the sub-areas \textit{Education} (one of the Framework-conditions for RTI), \textit{Tertiary Education} (a sub-area of the Core RTI System), and \textit{Digitization} (an RTI Cross-Cutting Issue)) are highlighted. This functionality is a bi-directional mapping based on the indicators analyzing these sub-areas.
\end{reqbox}
%\vspace*{5mm}

%%%%% was heißt "switching between different perspectives" im vorigen Absatz???
%%%% Die nächste Box ist in Überbleibsel, order?
%\vspace*{2.5 mm} 
%\begin{mdframed}[backgroundcolor=gray!20] 
%    {\bf Visualizing the RTI system with its areas:}\\       
%    These areas are interlinked as a color-coded graph which reflects that all areas are interconnected and maps their causal relations within the RTI system This affords comprehensive analyses at multiple information levels.
%\end{mdframed}

Evidently, the RTI monitor must provide a clear and comprehensive overview of the current status of the entire RTI key data to stakeholders. It must become clear to a stakeholder to what extent the strategic goals have already been achieved, and how the individual areas and sub-areas are performing. Therefore, a visualization concept has to be selected for the strategic goals that can illustrate the degree of target achievement. The performance of the areas and sub-areas is usually indicated in an international context, comparing them with those of other countries or regions. Given these interrelated visualizations between strategic goals and (sub)-areas as described above, a high achievement rate for strategic goals will be backed by high performance in the correlated (sub)-areas.

Further going into detail, the core of an RTI monitor is a comprehensive, data-driven, and, thus, evidence-based analysis of all identified sub-areas of the RTI system. 
A data-driven analysis must be based on a set of appropriate indicators describing the respective sub-area. The results for all the indicators must be visualized in a compact manner, while at the same time, stakeholders may access the original data sources for the indicators to ensure transparency. Even if the absolute values of the indicators deliver a first hint towards the performance of the RTI system in certain aspects, the complete picture is only provided if this performance is set in comparison with other RTI systems (e.g., of other countries or regions). 

Thus each indicator needs to be visualized as a relative value in percentage when compared to the one of other selected regions. Although the value of each indicator is essential, there should be an overall assessment value for the sub-area, calculated from all the individual indicators. Again, an easy-to-grasp visualization of this overall assessment value, in comparison to other RTI systems (of other countries and regions) is a key feature of an RTI monitor. Furthermore, it is important to show how the performance of the sub-area has evolved. In addition, to the visualization of the data-driven analysis of the sub-area, an interpretative textual analysis should be provided for the stakeholders. 

%\vspace*{5 mm}
\begin{reqbox}
 [Achievement of strategic goals and performance of (sub-)areas]
    To provide a quick overview of the status of strategic goal achievement the Austrian RTI monitor leverages so-called \textit{Harvey balls} (cf. Figure~\ref{fig:RTI-Level-1}), which are circular ideograms that are used to visualize the percentage of the achievement~\cite{Arnold2024}. This percentage is calculated based on a set of well-defined indicators. Austria's strategic goals~\cite{strat2030,circ}, based on national indicators determined by a special group of stakeholders of the federal ministries of Austria (i.e., Ministry of Labour and Economy, Ministry of Climate Action, Environment, Mobility, Innovation and Technology, Ministry of Education, Science and Research and the Federal Chancellery) of Austria, the so-called \textit{RTI Task Force}. FORWIT and WIFO work closely together with this task force to compute indicators to appropriately monitor the RTI goals. 
    
    %There is a separate page for each strategic goal, just as there is for the sub-areas of the RTI system, and a time series visualizes the current status quo. Additionally, we visualize \textit{how} the current status could develop in the future, to show the target achievement opportunity and anticipate the future (see Figure~\ref{fig:goal_ranking}).
    
    For visualizing the performance of the areas and subareas, the respective nodes are color-coded, reflecting an overall indicator presenting an average indicator based on all indicators of the respective sub-area~\cite{janger_2020,Arnold2024}. Note that a detailed discussion on the indicators of a sub-area is presented further below. The color coding resembles a traffic light system, based on a Likert scale (see Figure~\ref{fig:RTI-Level-1}). The colors green to light green indicate a performance of Austria that is at or above the average level of the respective Innovation Leaders. Yellow indicates a small gap to the average level, orange and red a large and a very large gap.    
\end{reqbox}
%\vspace*{5mm}

The analysis of a sub-area should be complemented by additional information relevant to the sub-area. Evidently, this includes documentation on the strategic goals supported by the corresponding sub-area. In addition, it is desired to establish links to further online information related to the sub-area. Such online information may be other dashboards and tools, as well as white papers and reports on the sub-area. To facilitate the navigation within the monitor, all other sub-areas related to the one under consideration should be easily accessible without going to the top page again.

%\vspace*{5 mm}
  The detailed analysis of a sub-area of the Austrian RTI monitor is exemplified by means of the sub-area ``Digitization'' in Figure \ref{fig:RTI-Level-2}. We have marked the different areas with boxes with dashed borders. In addition, we add a one-character identifier for the area in a grey circle.

In the context of an RTI monitoring system, indicators are metrics used to assess, track, and report on specific aspects of the RTI system. These indicators help in understanding performance, identifying trends, and making informed decisions. Evidently, the definition of appropriate indicators for each of the sub-areas is the most crucial task when developing an RTI monitor. It requires a thorough discussion with all the relevant stakeholders in the RTI system.

In order to be considered for an RTI monitor, an indicator must meet certain criteria: First and foremost, the indicator must be suitable to measure the sub-area under consideration and it must be relevant in the context of the strategic goals of the RTI system. Second, the indicator must not only exist in the local context of the RTI system but is a relevant indicator measured in other RTI systems, as well. This requirement for an indicator ensures its comparison with other countries and regions. Third, the data values for an indicator must be available and accessible in trusted information sources. In other words, the data collection module (cf. Figure~\ref{fig:architecture}) must be able to (semi-automatically) crawl the time series of the indicator from reliable external data sources. Please note that a prerequisite for this is open online access, which is unfortunately not always the case. Usually, the monitor depends also on non-disclosed data (e.g. from the OECD). Evidently, these cannot be used in their native form, but only data calculated based on them can be made available.

\begin{reqbox}
 [Detailed analysis of a sub-area]
    All relevant values and respective indicators relevant for the sub-area are visualized as bar charts in (cf. \textit{Area A} of the dashboard).Values are represented as percentages compared to the reference RTI system. with the endpoint colors, corresponding to the aforementioned color scheme. Note that in the Austrian RTI monitor one may dynamically select three alternative reference regions in \textit{Area J} on the upper right: (1) innovation leaders (the same for all areas and indicators), (2) Top 3 countries (may be different for each single indicator), and (3) EU average. Depending on the selected reference region, the values of the indicators and, thus, the bar chart will adapt dynamically. In \textit{Area B} stakeholders may access the original data on which the indicators are based upon.

    Based on these individual indicators for the sub-area, an overall assessment indicator for the sub-area is computed. The resulting overall indicator for the sub-area is shown in \textit{Area C} on the upper right. Evidently, the color code of this value corresponds to the one for the area on the top-level page (cf. Figure~\ref{fig:RTI-Level-1}) - so in both cases of the example it is ``orange'' for the sub-area ``Digitization''. Area C does not only show a single value for the overall indicator of the sub-area in the year of consideration (which may also be selected in \textit{Area J}, but shows the evolvement of the indicators over the years.

    A textual interpretation of the data of the sub-area is provided in \textit{Area E} on the lower right. \textit{Area F} further shows the strategic goals (cf. Figure \ref{fig:RTI-Level-1}) related to the selected sub-area. In addition, \textit{Area F} lists also another set of goals, which are additional impact targets of the Austrian ministries. \textit{Area G} provides links to other relevant tools, which are in this sub-area (``Digitization'') related to the European Green Deal and the Sustainable Development Goals. Furthermore, \textit{Area H} provides access to different types of documents (e.g., recommendations, statements, studies) created on behalf of the Austrian Council (i.e., FORWIT). \textit{Area I} provides short-cuts to all other sub-areas that are related to the one under consideration, in this case, related to``Digitization''.    
\end{reqbox}

There are two main types of indicators: single indicators and composite indicators. A single indicator is a metric that measures one specific attribute or variable. It provides direct, straightforward information about a particular aspect of the system being monitored. Single indicators are typically simple to calculate and interpret. A composite indicator, on the other hand, is a metric that combines multiple individual indicators to provide a more comprehensive assessment. They are often created by aggregating, averaging, or applying a formula to several single indicators.
\begin{figure*}[th!]
\centering
\begin{subfigure}{0.475\textwidth}
\frame{
 \includegraphics[width=1\linewidth]{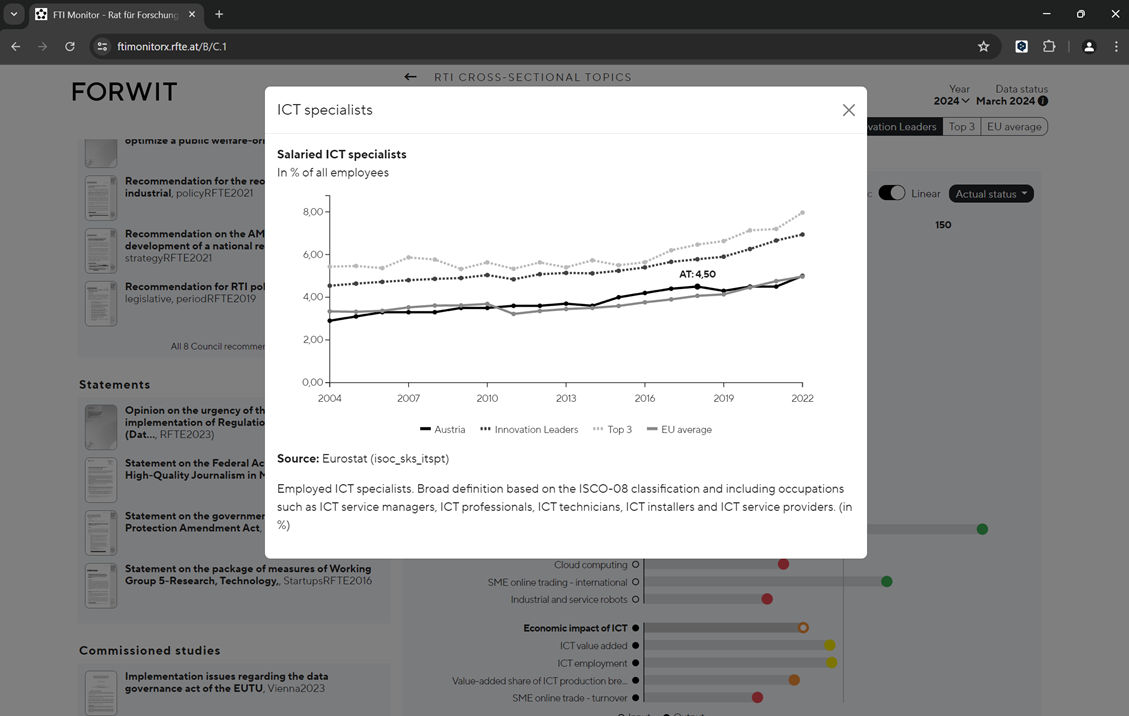}
 }
    \caption{Sub-area Digitization: Time series of the Input Indicator analyzing ICT specialists.}
    \label{fig:timeseries_ICT}

\end{subfigure}
\hfill
\begin{subfigure}{0.50\textwidth}
       \centering
       \frame{
    \includegraphics[width=1\linewidth]{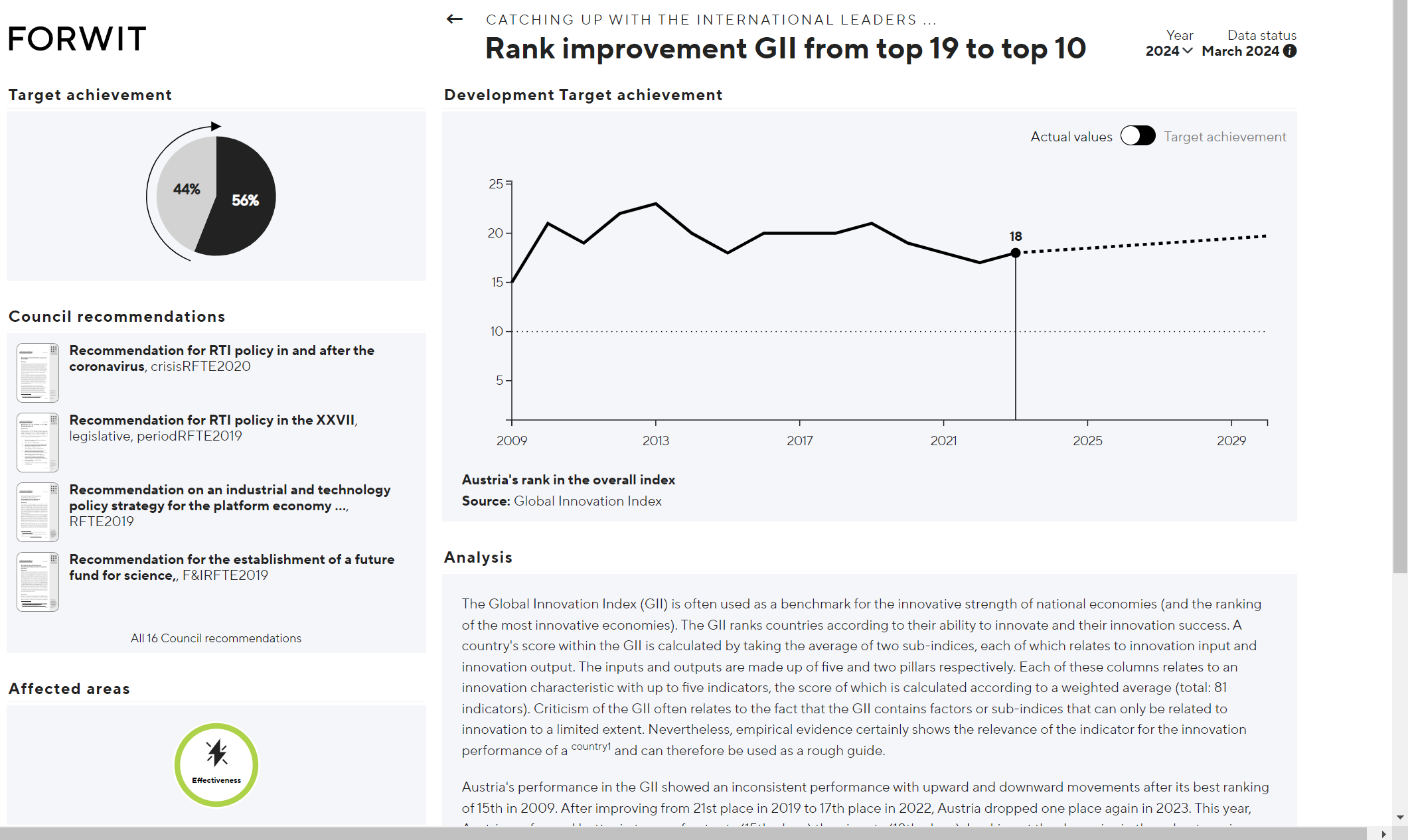}
    }
    \caption{Strategic RTI goal -- Rank improvement of the Global Innovation Index.}
    \label{fig:goal_ranking}
\end{subfigure}        
\caption{Examples for Areas and Goals from the FORWIT Monitor}
\label{fig:figures}
\end{figure*}

Independent of being a single or composite indicator, each indicator may be an input or output indicator. An \textit{input indicator} measures the resources, efforts, and investments put into the research, technology, and innovation activities. These indicators are concerned with the "inputs" that contribute to the process of generating innovations. An \textit{output indicator} measures the immediate results or products of the research, technology, and innovation activities. These indicators focus on the "outputs" that are directly produced from the inputs and processes.

   The predecessor of the Austrian RTI Monitor was a printed report on the Austrian RTI system. For this printed report, the Austrian Institute of Economic Research (WIFO) developed a 
   
   %\vspace*{5 mm}
\begin{reqbox}
 [Indicators of the Austrian RTI Monitor]

    All indicators are collected from national and international sources, like OECD, Eurostat, Invest Europe, World Bank, THE, Statistic Austria, and others. The Austrian RTI Monitor always provides a link to the source of origin and a brief textual description of the indicator. Needless to say, to search various sources and semi-automatically collect data (only for those data sources providing open access) a data collection module and a database for storage are required as depicted in Figure~\ref{fig:architecture}. Both components are out of the scope of this paper but will be discussed in follow-up work. 

    The indicators defined for a sub-area appear in \textit{Area A} of the dashboard, and the overall indicator -- presenting an average indicator of all indicators -- in \textit{Area C} (cf. Figure \ref{fig:RTI-Level-2}). The indicators analyzing a sub-area are grouped   into input and output indicators, as well as single and composite indicators. Composite indicators appear in bold font and list the included single indicators.  Input indicators show an unfilled circle after their name, whereas output indicators come with a filled circle (see \textit{Area A} of Figure~\ref{fig:RTI-Level-2}). All of these indicators are visualized as color-coded bar charts. 
    
    The color code depends on the data value to a specific point in time (the so-called data cut-off) and the selected reference RTI system in \textit{Area J}. Since a data value at a specific point in time represents a ``snapshot'', an additional time series visualization  provides stakeholders with a historical analysis. Figure~\ref{fig:timeseries_ICT} depicts an example of the time series for the  ICT specialists indicator, an input indicator of the sub-area \textit{Digitization}). The time series always shows the evolution of the indicator for (1) Austria, (2) the Innovation Leaders, (3) the Top 3 Countries of this indicator, and (4) the EU average. Hovering over the time series, e.g. Austria (AT), further reveals the value of this indicator for a specific year (see Figure~\ref{fig:timeseries_ICT} AT: 4,50 in 2018). Additionally, the dashboard module of the RTI Monitor offers the functionality to display the change contributions (positive or negative) of each indicator to the overall indicator. 
\end{reqbox}

   benchmarking tool with already well-defined input- and output-indicators~\cite{janger_2020}. This set of indicators provided a very reasonable starting point for the Austrian RTI Monitor. Updates to these indicators are always suggested by the Austrian Research, Sciences, Innovation and Technology Council after discussion with the WIFO and considering the feedback of all relevant stakeholders in the Austrian RTI system.
   
\vspace*{2.5 mm}
\begin{reqbox}
 [Strategic Goals Performance]
As already mentioned, the Austrian RTI Monitor uses Harvey balls to visualize the current status of the strategic RTI goals as well as the goals of the strategic circular economy initiative~\cite{strat2030,circ} to a specific point in time~\cite{Arnold2024}. In contrast to the benchmarking of the sub-areas of the RTI system, Austria’s strategic goals are based on national indicators determined by the federal ministries of Austria and the RTI task force. The user can see at a glance to what percentage the goal has already been achieved. Figure~\ref{fig:goal_ranking} shows the goal of “Rank improvement GII from 19 to top 10” which has already been achieved to 56\% (see the Harvey ball in the upper left area ). On the right-hand side of the figure, a time series is displayed. Unlike to the time series shown when clicking on an indicator in one of the sub-areas, this time series shows the ranking of the indicator to specific points in time, and in addition, provides an outlook on future target achievement from the current point in time. Figure~\ref{fig:goal_ranking} displays that Austria is ranked 18th and will not be able to improve this position significantly from today's point of view.

\end{reqbox}

So far, we have mainly concentrated on the analysis of the indicators of a sub-area. In addition, the performance with respect to the strategic goals is usually of great interest to the stakeholders of the RTI community. Accordingly, visualizing the respective indicators as a time series is another key feature of an RTI monitor. Stakeholders usually want to know whether a strategic target has already been achieved or, if not, whether it can still be met in the (near) future. Furthermore, stakeholders are interested in the rank of a country in an international comparison based on the underlying indicator.

% \begin{figure}[htb]
%     \centering
%     \includegraphics[width=1\linewidth]{figures/timeseries.png}
%     \caption{Sub-area Digitization: Time series of the Input Indicator analyzing ICT specialists.}
%     \label{fig:timeseries_ICT}
% \end{figure}

% \begin{figure}[t]
%     \centering
%     \includegraphics[width=1\linewidth]{figures/Goal_GII.png}
%     \caption{Strategic RTI goal by the example of the Rank improvement of the Global Innovation Index.}
%     \label{fig:goal_ranking}
% \end{figure}

% Der Absatz wurde bereits writer oben behandelt.
%In addition to the benchmarking within the RTI system, there is also a benchmarking needed for monitoring the strategic goals in a structured manner. Therefore, the requirements for visualizing these goals are, on the one hand, to give a detailed overview of them, and on the other hand, to visualize their actual status quo. 

A further requirement on the dashboard is to provide users free access to all available documents interesting for a specific area.  The aim is to open such documents with a simple mouse click and download them as needed. Additionally, in all specific areas, it should be possible to link to other sources or dashboards (e.g., the Sustainable Development Goals dashboard provided by Eurostat, etc.) to avoid unnecessary information duplication.

\vspace*{2.5 mm}
\begin{reqbox}
 [Display certain documents]
To fulfill this requirement, the User Management Modul of the Austrian RTI Monitor does not restrict the access to the Dashboard module, thereby guaranteeing open access to all interested stakeholders. In addition, all documents (e.g., studies, recommendations, statements, etc.) that have been prepared on behalf of the Austrian Council for Sciences, Innovation and Technology (FORWIT) are assigned to the corresponding areas of the RTI system and are offered for free download. For example, in the sub-area \textit{Digitization}, the user can visualize each recommendation or statement to this specific subject made by FORWIT, or studies on this topic commissioned by FORWIT (see \textit{Area H} of Figure~\ref{fig:RTI-Level-2}). Furthermore, links to other sources such as the European Green Deal and the Sustainable Development Goals (SDGs) are covered in \textit{Area G}.

\end{reqbox}

In the next section, we provide an insight into "how" a negotiation process among different stakeholders can be set up using a framework established in the Software Engineering domain. We will again use our own work on the RTI Monitor as an example.

\section{Conclusion and Outlook}
\label{sec:conclusion}

This paper focuses on the requirements for developing a Web-based monitoring tool within the RTI environment. For this purpose, we outlined a three-tier software architecture for realizing such a tool. The presented architecture comprises several well-defined modules (cf. Sec.~\ref{sec:Components}, Fig.~\ref{fig:architecture}): On the bottom tier, the Data Collection Module accesses relevant data from various trusted data sources. The Data Integration Module ensures that the data is integrated regardless of format to the Data Store Module (i.e., the database). At the middle layer, the so-called business tier, the Analytics Engine accesses this database to perform relevant analytics for decision-making. In addition, an optional Notification Module may proactively inform stakeholders about relevant updates to the system. A Configuration Module is an advanced module that enables on-the-fly configurations of the system without the need to hard-code them.  The top tier represents the user interfaces. Regular users interact via the Dashboard Module that provides intuitive visualizations, e.g., those of the indicators. Administrators use the Admin Module to specify the configurations of the system.

Since the Dashboard Module serves as the interaction interface from the data to the users of the monitoring tool, the functionalities provided by this module are crucial for the acceptance of the tool. Therefore, we concentrated in this paper on the requirements of the dashboard module describing what the stakeholders may expect from this interaction interface. For a better understanding of these requirements, we discuss how these requirements were met in the development of the Austrian RTI monitor. In particular, we defined what concepts should be covered by a dashboard module and how these concepts were customized for the Austrian RTI monitor.

Capturing the requirements of a system to be developed requires also the commitment of the stakeholders of the application domain. Therefore, the process of eliciting those requirements has to follow a structured process to ensure that the expectations of all stakeholders using the system are met. In this paper, we suggested applying a Viewpoint-guided Requirements Engineering process based on the VORD-Model (cf. Sec.~\ref{sec:RE_Process}) for this purpose and illustrated how this requirements engineering process may be used in the context of an RTI monitor.

While this paper has extensively covered the requirements from the users' perspective, our future work will introduce a comprehensive meta-model that presents these requirements from a technical point of view. This meta-model will aim to bridge the gap between user expectations and technical specifications, ensuring a more integrated and efficient system development process.

%The paper highlights the advantages of such an on-line monitoring tool in policy-making environments. It promises a real-time data analysis within the RTI framework, promising a deeper insight into how various factors affect innovation performance. Key features include a comprehensive analysis of the system's strengths and weaknesses, evaluations of historical data by time series, and comparisons between different innovation ecosystems.

%  \section*{Acknowledgment}
% Txxx
 
\balance

\bibliographystyle{abbrv}
\bibliography{refs} 

\end{document}